\documentclass[doublecol]{epl2}

\usepackage{graphicx}
\usepackage{graphics}
\usepackage{latexsym}
\usepackage{subfigure}
\usepackage{multirow}
\usepackage{amsmath}
\usepackage{amssymb}

\newcommand{\be}{\begin{equation}}
\newcommand{\ee}{\end{equation}}
\newcommand{\bea}{\begin{eqnarray}}
\newcommand{\eea}{\end{eqnarray}}

\title{Universal scaling dynamics in a perturbed granular gas}

\author{Zahera Jabeen \and R. Rajesh \and Purusattam Ray}
\institute{
Institute of Mathematical Sciences, CIT Campus, Taramani, Chennai-600113,
}

  
\date{\today}

\abstract{
We study the response of a granular system at rest to an instantaneous 
input of energy in a localised region. We present scaling arguments that 
show that, in $d$ dimensions, the radius of the resulting disturbance 
increases with time 
$t$ as $t^{\alpha}$, and the energy decreases as $t^{-\alpha d}$, where 
the exponent $\alpha=1/(d+1)$ is independent of the coefficient of 
restitution.  We support our arguments with an exact calculation in one 
dimension and event driven molecular dynamic simulations of hard sphere 
particles in two and three dimensions.
}

\begin{document}
\maketitle

Granular systems, predominantly characterized by dissipative collisional 
dynamics, are ubiquitous in nature and exhibit a wide variety of very 
rich and striking physical phenomena \cite{jaeger1996}. Although many 
experimental studies have captured the complexity of these systems by 
studying phenomena ranging from clustering instability, co-existence of 
phases to non-Maxwellian velocity distributions (see 
\cite{jaeger1996,kudrolli2004} for reviews), the theoretical 
understanding of these systems is far from complete (see 
\cite{aranson2006,kadanoff1999,brilliantovbook} for reviews). Hence, it 
is imperative to study simple models that capture some distinctive 
features of the system, yet are amenable to analysis.

A model that has attracted considerable attention in the past is the 
freely cooling granular gas, where the particles move ballistically and 
lose energy only through inelastic collisions 
\cite{carnevale,goldhirsch1993,trizac1,britoernst,frachebourg,bennaim1999,nie2002,trizac2,puri,cattuto,shinde2007,shinde2009}. 
Starting from a homogeneous spatial distribution of particles with 
velocities drawn from a normalizable distribution function, simulation 
studies show that after an initial regime when energy decays as $E_t\sim 
t^{-2}$ (Haffs law)\cite{haff}, clustering instability sets in 
\cite{goldhirsch1993}. The long time behavior of the system is 
universal: the energy decays algebraically with an exponent which 
depends on the dimension but not upon the coefficient of restitution 
\cite{bennaim1999,nie2002,trizac2,shinde2009}.  The exponent is known 
analytically in one dimension through a mapping to the Burgers equation 
($E_t\sim t^{-2/3}$) \cite{kida,frachebourg}. In higher dimensions, the 
exponents obtained from the analogy to Burgers equation ($E_t\sim 
t^{-d/2},d\geq 2$) \cite{nie2002} differ from that obtained from mean 
field scaling arguments ($E_t\sim t^{-2d/{d+2}}$)\cite{carnevale} and 
from simulations of the Boltzmann equation \cite{trizac4,trizac2}, 
leading to an uncertainty in the precise value of the exponents in two and 
higher dimensions.

\begin{figure} 
\includegraphics[width=4cm]{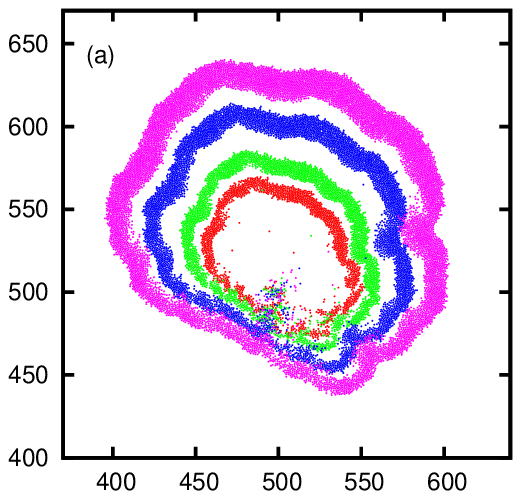} 
\includegraphics[width=4.2cm]{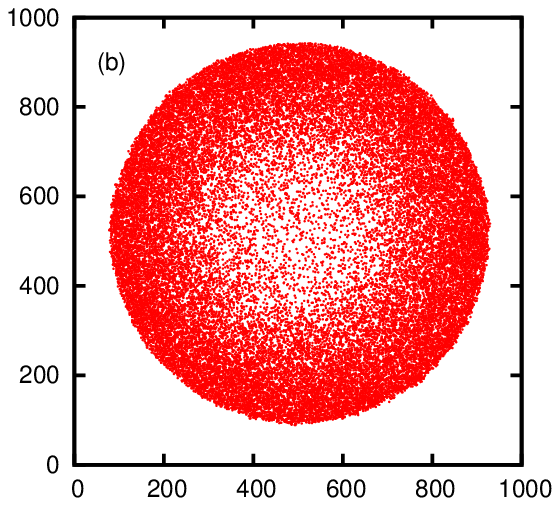} 
\caption{(Color online) Shown 
are the positions of particles that have undergone at least one 
collision, following input of energy at $(500,500)$ for (a) the 
inelastic case ($r=0.1$) at times $t=5000, 10000, 20000, 50000$, and (b) 
elastic case ($r=1.0$) at time $t=25000$.} \label{mfront} 
\end{figure}

In this paper, we consider a simple and tractable model of a cooling 
granular gas where the particles are initially at rest and the system is 
perturbed by imparting momentum to a single particle. This in turn leads 
to motion of other particles by inter-particle inelastic collisions, and 
the particles cluster to form a nearly spherical shell that propagates 
radially outwards in time [see Fig.~\ref{mfront} (a)]. Using scaling 
arguments and numerical simulations, we show that the scaling behaviour 
of energy and the radius of the disturbance with time is independent of 
the coefficient of restitution. The results obtained from scaling 
arguments are confirmed by an exact calculation in one dimension and 
event driven molecular dynamics simulations in two and three dimensions.

The corresponding problem when collisions are elastic is the classic 
Taylor-von Neumann-Sedov problem of shock propagation following a 
localized intense explosion \cite{sedov}. In this case, the particles 
remain homogeneously distributed [see Fig.~\ref{mfront} (b)] and the 
exponents can be obtained by simple dimensional analysis \cite{taylor}, 
while the scaling functions can be calculated exactly following a more 
detailed analysis \cite{sedov,neumann}. The simulations and scaling 
arguments for a hard sphere model with elastic collisions were recently 
done in Ref. \cite{antal2008}. Signal propagation has also been studied in
excited dilute granular gas \cite{losert} as well as in dense static granular
material (see \cite{luding} and references within).

Our model consists of a collection of monodisperse hard spheres (in 
simulation we have taken $2.5 \times 10^5$ and $2\times 10^6$ particles 
in two and three dimensions respectively) of finite diameter (unity in 
simulation) distributed randomly in space such that no two particles overlap
(in simulation the number 
density $n=0.25$ in both two and three dimensions). Periodic boundary
conditions are implemented in all directions. All the particles 
are initially at rest. A single particle is chosen at random and given a 
velocity of unit magnitude along a random direction. The particle motion 
is ballistic till it collides with other particles. The collisions 
conserve momentum and the velocities change deterministically according 
to the following collision rules: if the velocities before and after 
collision are ${\bf u}_1$, ${\bf u}_2$, and ${\bf v}_1$, ${\bf v}_2$ 
respectively, then \be {\bf v}_{1,2}={\bf u}_{1,2}-\epsilon [{\bf 
n}.({\bf u}_{1,2} -{\bf u}_{2,1})] {\bf n}, \ee where $r=2 \epsilon-1 
(0<r<1)$ is the coefficient of restitution and ${\bf n}$ is the unit 
vector directed from center of particle $1$ to center of particle $2$. 
Thus, the tangential component of the relative velocity remains 
unchanged, while the magnitude of the longitudinal component is reduced 
by a factor $r$.

For $r<1$, the system undergoes inelastic collapse in which infinite 
collisions take place in finite time \cite{Mcnamara}. This computational 
difficulty is avoided by making the collisions elastic when the 
longitudinal relative velocity is less than a cutoff velocity $\delta$ 
\cite{bennaim1999}. This qualitatively captures the experimental 
situation where $r$ is seen to be a function of the relative velocity 
\cite{raman,fauve}. In our simulation, we set $\delta=10^{-4}$.

Consider now the result of a typical simulation [see 
Fig.~\ref{mfront}(a)].  Let $R_t$ be the typical radius of the shock 
profile, $v_t$ the typical speed, $N_t$ the number of active particles 
(particles that have undergone collisions), and $E_t$ the total kinetic 
energy at time $t$.  These quantities are related to each other through 
simple scaling relations. The speed $v_t$ is related to $R_t$ as $ 
v_t\sim dR_t/dt$. The number of particles that have undergone collisions 
is proportional to the volume swept out by the disturbance: $ N_t \sim 
R_t^d$, where $d$ is the dimension. Energy is then given by $E_t \sim 
N_t v_t^2$.

We look for scaling solutions of the kind $R_t \sim t^\alpha$, 
where $\alpha$ is a scaling exponent. 
Then, 
\bea
v_t & \sim & t^{\alpha-1}, \label{eq:vel}\\
N_t &\sim & t^{\alpha d}, \label{eq:number}\\
E_t &\sim & t^{\alpha d + 2 \alpha -2}. \label{eq:energy}
\eea

The above relations hold good for both elastic and inelastic collisions. We 
now analyze the two cases separately. For the elastic gas, energy is a 
constant of motion. This implies
\be
\alpha=\frac{2}{d+2}, \quad r=1,
\ee
coinciding with the results for one and two dimensions in Ref. 
\cite{antal2008} and for three dimensions in Ref. \cite{taylor}.

For the inelastic case, there is one unknown exponent $\alpha$ which is 
determined by the following argument. A short time after the initial 
perturbation, the particles that have undergone at least one collision 
concentrate themselves into a narrow band. Though the data shown in 
Fig.~\ref{mfront}(a) is for $r=0.1$, clustering is seen for all $r<1$. 
Due to this spatial structure, there is no radial momentum transferred 
from particles at a certain angle to those that are diametrically 
opposite, or in other words, the radial momentum is conserved. The 
radial momentum carried by the particles in a small solid angle 
$d\Omega$ scales as $v_t R_t^d d \Omega$.  The conservation law implies 
that $v_t R_t^d \sim \mathrm{const}$, or equivalently, $v_t \sim 
{R_t}^{-d} \sim t^{-\alpha d}$. Comparing with Eq.~(\ref{eq:vel}), we 
immediately obtain
\be 
\alpha=\frac{1}{d+1}, \quad r<1. 
\label{eq:inelastic} 
\ee

In one dimension, the above scaling result can be checked by a simple 
calculation. Consider the sticky limit $r=0$, when the particles 
coalesce on collision. Let particles of unit mass be initially placed on 
a lattice with spacing $a$. Let the particle at the origin be given a 
velocity $v_0$ to the right. When this particle collides with its 
neighbor, it coalesces with it. The mass of this composite particle 
after $m$ collisions is then $m$, and its velocity, given by momentum 
conservation, is $v_m=v_0/m$ towards the right. The time taken for $m$ 
collisions is given by
\bea
t_m&=& \sum_{i=0}^{m-1} \frac{a}{v_i},\\
&=&  \frac{a m(m-1)}{2 v_0}.
\eea
At large times, $m \approx \sqrt{2 v_0 t/a}$. But $m$ is identical to 
$N_t$ and $R_t$, which by definition scales as $t^\alpha$. This gives 
$\alpha=1/2$, consistent with that obtained by setting $d=1$ in 
Eq.~(\ref{eq:inelastic}).
\begin{figure}
\includegraphics[width=\columnwidth]{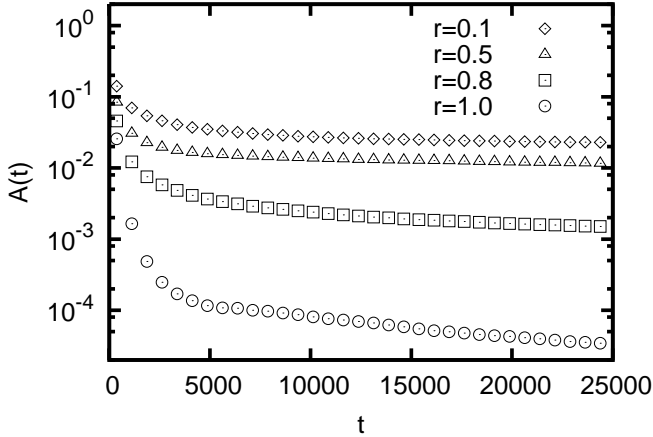}
\caption{The anisotropy index $A(t)$ of the band in two dimensions is plotted as a function of time $t$ for different
values of the coefficient of restitution $r$. $A(t)$ converges to a value
less than one for all $r$. 
\label{aindex}} 
\end{figure}

In two and three dimensions, the scaling arguments are tested 
numerically using event driven molecular dynamics simulations 
\cite{rapaportbook}. The data presented is averaged typically over $100$ 
different initial realizations of the particles. All lengths are 
measured in units of the particle diameter, and time in units of initial 
mean collision time $1/(v_0 n^{1/d})$, where $v_0$ is unity in the
simulations.
We first check the validity of the assumption of a 
single length scale $R_t$. Fig.~\ref{aindex} shows the variation in two dimensions of the 
anisotropy index $A(t)$ with time, where the anisotropy index is given by $A(t)= \langle 
[(\lambda_1-\lambda_2)/(\lambda_1+\lambda_2)]^2 \rangle$, 
$\lambda_1, \lambda_2$ being the eigenvalues of the moment of inertia 
tensor \cite{momentofinertia}. If the transverse and 
longitudinal radii scale differently with time, then $A(t)$ should 
converge to unity at large times. However, $A(t)$ is found to converge to 
a constant less than one for all $r$. For $r=1$, $A(t)$ converges to 
zero at large times. We conclude that though the shape of the front is 
anisotropic for $r<1$, all length scales scale identically with time.

\begin{figure} 
\includegraphics[width=\columnwidth]{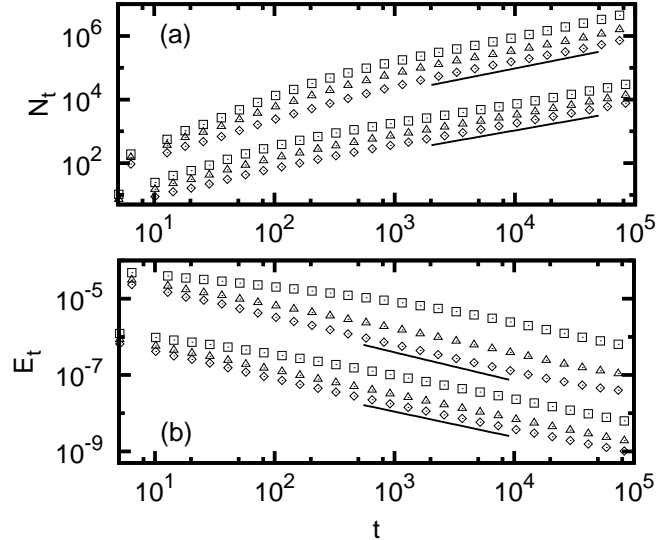} 
\caption{Simulation results for the (a) the mean number of active 
particles $\langle N_t \rangle$ and (b) the mean kinetic energy $\langle 
E_t \rangle $ as a function of time $t$. In both the plots, the top 
three curves correspond to three dimensions and the bottom three curves 
correspond to two dimensions.  The different data correspond to the 
coefficients of restitution $r=0.1 (\Diamond), 0.5(\triangle), 
0.8(\square)$. The solid lines have exponents obtained from scaling 
theory. The data have been shifted for the sake of clarity.\label{ntet}} 
\end{figure}

We check the scaling relations Eqs.~(\ref{eq:number}), 
(\ref{eq:energy}), and (\ref{eq:inelastic}) by measuring the mean number 
of active particles $\langle N_t \rangle$ and the mean total kinetic 
energy $\langle E_t \rangle$ as a function of time. In two dimensions, 
the scaling argument gives $\langle N_t \rangle\sim t^{2/3}$, $\langle 
E_t \rangle \sim t^{-2/3}$, while in three dimensions $\langle N_t 
\rangle \sim t^{3/4}$, $\langle E_t \rangle \sim t^{-3/4}$. In 
Fig.~\ref{ntet}(a) and (b), we show the variation with time of $N_t$ and 
$E_t$ in two and three dimensions. For larger $r$, it takes longer time 
to reach the scaling regime. This crossover time $t_c^{(1)}$ reflects 
the transition of the particles from the initial homogeneous spatial distribution  to the clustered state. We find that $t_c^{(1)}$ diverges in 
the elastic limit as $t_c^{(1)}\sim(1-r^2)^{-\phi_1}$, where $\phi_1 
\approx 2.25$ in two dimensions and $\phi_1 \approx 3.0$ in three 
dimensions. In addition, at large times, the system crosses over to the elastic 
regime when $v_t \sim \delta$. This crossover time scales as 
$t_c^{(2)}\sim \delta^{-\phi_2}$ where $\phi_2= 1/(1-\alpha)$ [$3/2$ in 
$d=2$ and $4/3$ in $d=3$].  Within these limitations, the numerical data 
shows good agreement with the theoretical prediction shown with solid 
lines.

\begin{figure}
\includegraphics[width=\columnwidth]{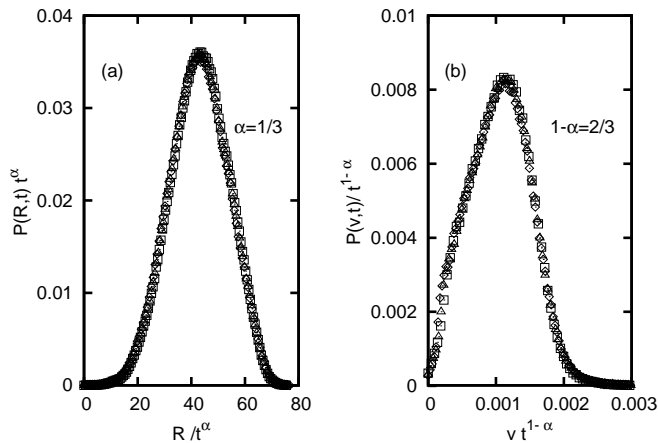}
\caption{Results in two dimensions for (a) the radial distribution 
function $P(R,t)$ and (b) the velocity distribution function $P(v,t)$, 
when scaled as in Eqs.~(\ref{prt_eq}) and (\ref{pvt_eq}) with scaling 
exponent $\alpha=1/3$. The scaling collapse has been obtained for times 
$t=25000 (\Diamond), 37500 (\triangle),$ and $50000 (\square)$. The 
coefficient of restitution is $r=0.1$ \label{2d_dist}}
\end{figure}

\begin{figure}
\includegraphics[width=\columnwidth]{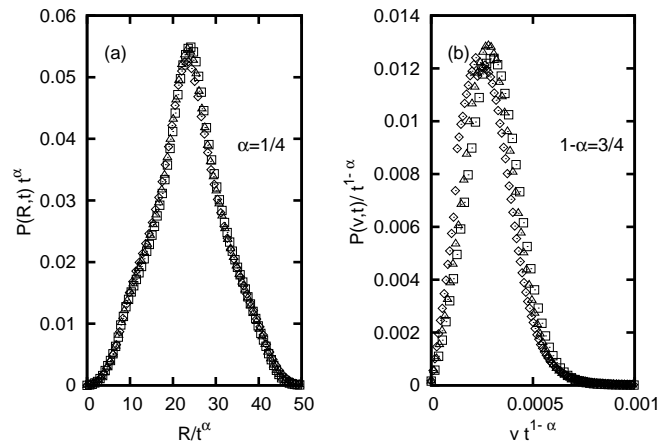}
\caption{Results in three dimensions for (a) the radial distribution 
function $P(R,t)$ and (b) the velocity distribution function $P(v,t)$, 
when scaled as in Eqs.~(\ref{prt_eq}) and (\ref{pvt_eq}) with scaling 
exponent $\alpha=1/4$. The scaling collapse has been obtained for times 
$t=30000 (\Diamond), 50000(\triangle)$ and $75000(\square)$.  The 
coefficient of restitution is $r=0.1$. \label{3d_dist}}
\end{figure}

We check the scaling relations for $R_t$ and $v_t$ by studying the 
radial and the velocity distribution function. The radial distribution 
function $P(R,t)$ measures the mean number of active particles at a 
distance $R$ from the center of mass of the active particles at time 
$t$. The velocity distribution function $P(v,t)$ measures the 
probability that a randomly chosen active particle has speed $v$ at time 
$t$. These distribution functions should be a function of a single 
scaling variable:
\bea
P(R,t) &= & t^{-\alpha} f_1(R t^{-\alpha}), \label{prt_eq} \\
P(v,t) &=& t^{1-\alpha} f_2(v t^{1-\alpha}), \label{pvt_eq}
\eea
where $f_1$ and $f_2$ are scaling functions. These scaling collapses are 
verified numerically in two dimensions (see Fig.~\ref{2d_dist}) and in 
three dimensions (see Fig.~\ref{3d_dist}). The data shown is for one 
value of the coefficient of restitution ($r=0.1$), but the same is 
observed for other values of $r$. The scaling function $f_2(v 
t^{1-\alpha})$ decays exponentially at large speeds $v$. Such 
non-Maxwellian behaviour is typical of granular systems 
\cite{menon,kudrolli,ernst}. We also observe that the faster particles 
are in the inside edge of the collapsed band, thus making the 
bands stable.

We also studied the structure of the collapsed bands. For that, the 
packing fraction of the particles in the bands was numerically 
calculated by dividing the space into cells of linear length $10$, and 
counting the number of particles in each cell. For all $r<1$, the 
typical packing fraction seen at large times ranges from $0.78-0.82$ in 
two dimensions. This value is very close to 
$0.84$, the packing fraction of random close
packed structures seen in jamming of frictionless spherical particles
\cite{williams}.
For $r=1$, the 
packing fraction is $\sim 0.47$, showing that the particles are very 
loosely packed.

To conclude, we studied the problem of shock propagation in granular 
(inelastic) systems and obtained scaling solutions for the problem. In 
one dimension, the exact result for the sticky limit ($r=0$) 
corroborated the scaling solution. In two and three dimensions, we 
verified our results using event driven molecular dynamics simulations. 
Our analysis showed conclusively the universality ($r$-independence) of 
the scaling solutions and its dependence only on the spatial dimension. 
We retrieved the earlier results for the classic Taylor-von 
Neumann-Sedov problem corresponding to the elastic limit ($r=1$). For 
$r<1$, we obtained an explicit expression for the scaling exponent in the 
late time cooling which has hitherto remained inconclusive for the 
related problem of the freely cooling granular gas.

The model discussed in this paper also has experimental significance. 
Direct experiments on freely cooling gas are difficult due to friction 
and boundary effects. Recent experiments reproduced the energy decay law in
the homogeneous cooling 
regime \cite{maa}, but not in the clustered regime. The boundary effects 
will be eliminated if the granular gas is initially at rest, making the 
problem discussed in this paper more easily reproducible in the laboratory.

%\bibliography{refn}

\end{document}